\documentclass[a4paper, amsfonts, amssymb, amsmath, reprint, showkeys, nofootinbib, twoside, onecolumn,notitlepage]{revtex4-1}
\def\prl{{Phys.~Rev.~Lett.}}    % Physical Review Letters
\usepackage{amsmath,amstext}
\usepackage[T1]{fontenc}
\usepackage{amssymb}
\usepackage{graphicx}
\usepackage{ae,aecompl}

\DeclareFontFamily{OT1}{pzc}{}
\DeclareFontShape{OT1}{pzc}{m}{it}{<-> s * [1.10] pzcmi7t}{}
\DeclareMathAlphabet{\mathpzc}{OT1}{pzc}{m}{it}

\usepackage{hyperref}
\usepackage{amsmath}
\usepackage{amssymb}
\usepackage{mathtools}
\usepackage{bm}
\usepackage{cleveref}
\usepackage{tensor}
\usepackage{braket}
\usepackage{enumitem}
\usepackage{mhchem}
\usepackage{amsthm}
\usepackage{nccmath}
\usepackage{mathrsfs}
\usepackage{color}

\newcommand{\spc}{\quad \quad \quad}

\newcommand{\K}{{\mathcal{K}}}

\def\be{\begin{equation}}
\def\ee{\end{equation}}
\def\beq{\begin{eqnarray}}
\def\eeq{\end{eqnarray}}

\theoremstyle{definition}

\theoremstyle{theorem}

\begin{document}
\title{Spontaneous charge separation in accelerating relativistic plasmas}
\author{L.~Gavassino}
\affiliation{Department of Mathematics, Vanderbilt University, Nashville, TN, USA}

\begin{abstract}
The Stewart-Tolman effect posits that accelerating conductors exhibit both charge separation and rest-frame electric fields (``inertia of charge''), while the Ehrenfest-Tolman effect
states that acceleration induces temperature gradients  (``inertia of heat''). We study the interplay of these effects in thermodynamic equilibrium. Specifically, we derive from first principles a partial differential equation governing the electrothermal stratification of a fully ionized plasma in equilibrium under irrotational relativistic accelerations in curved spacetime. We then solve it in two settings: a plasma enclosed in a uniformly accelerated box, and a plasma shell suspended above a black hole horizon. The resulting electric fields are found not to depend on the electric conductivity of the medium.
\end{abstract} 

\maketitle
\section{Introduction}
\vspace{-0.3cm}

Recent studies have significantly reshaped our understanding of relativistic thermodynamics, leading to unexpected and practically valuable discoveries \cite{Banerjee:2012iz,BecattiniQuantumCorrections2015,BecattiniBeta2016,Becattini:2019poj,Becattini:2020qol,Palermo:2021hlf,GavassinoTermometri,GavassinoGibbs2021,GavassinoCausality2021,GavassinoSymmetric2022nff,MullinsInfo2023tjg,GavassinoUniveraalityI2023odx,Mullins:2025vqa,Pei2025}. This progress has encouraged a return to unaddressed foundational questions, which can now be approached with greater theoretical precision and rigor. Here, we investigate one such question: What is the equilibrium configuration of a relativistic plasma column undergoing constant acceleration?

At first, the strategy for determining the answer seems straightforward: Just solve the relativistic MagnetoHydroDynamic (MHD) equations \cite{Mizuno:2024tis} with the relevant boundary conditions (e.g., a moving ``floor'' that pushes the plasma). However, upon closer examination, one realizes that this approach is not guaranteed to work. First of all, it is not obvious that the resulting configuration will be a genuine maximum entropy state. More importantly, one should recall that, in its standard formulation, MHD breaks down at large accelerations. In fact, MHD rests on the assumption that, in the fluid's local rest frame, charge neutrality is effectively maintained, and no electric fields can be sustained without Ohmic dissipation. Unfortunately, these assumptions are known to break down in accelerating conductors, as demonstrated in a landmark experiment by Tolman and Stewart \cite{TolmanStewart1916,WangStewartTolman2023,BeiPhD}.

\begin{figure}
    \centering
\includegraphics[width=0.47\linewidth]{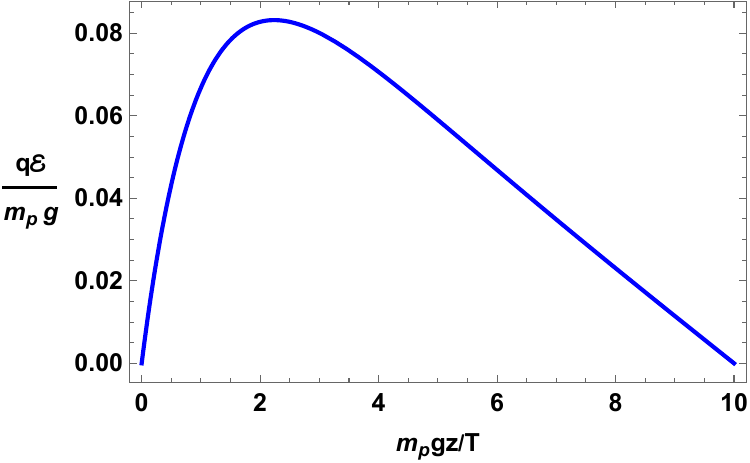}
\includegraphics[width=0.47\linewidth]{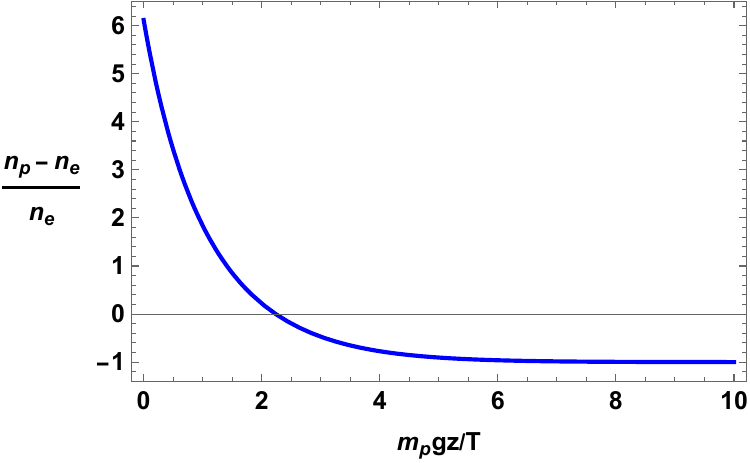}
\caption{Equilibrium electric field $\mathcal{E}$ (left panel) and charge separation $(n_p{-}n_e)/n_e$ (right panel) in a column of fully ionized hydrogen plasma undergoing acceleration $g$. This configuration solves the system \eqref{NewtonianStrat}-\eqref{Coulomblaw} with $\mathcal{Z}_p=\mathcal{Z}_e=20 q^2 T/(m_p^2 g^2)$, for a plasma confined in a region with height $10T/(m_p g)$. The boundary conditions are global charge neutrality plus the absence of externally imposed electric fields, which gives $(\partial_z\varphi)_{\text{Boundary}}{=}0$. The acceleration number for this configuration is $\mathscr{A}\sim 5$, so MHD does not apply (see footnote \ref{footononon1}).}
    \label{fig:Newtonian}
\end{figure}

If the system is non-relativistic, plasma physicists know how to improve upon the MHD prediction \cite[\S 1.6]{bellan_2006}. Thanks to the equivalence principle, a plasma subject to boundary conditions undergoing uniform acceleration $(0,0,g)$ is equivalent to a plasma with static boundary conditions, under the influence of a gravitational potential $\Phi \,{=}\, gz$. According to statistical mechanics, the equilibrium number densities of protons ($n_p$) and electrons ($n_e$) must then follow Boltzmann distributions that account for both gravitational and electrostatic potential energy \cite[\S 4.2]{huang_book}:
\begin{equation}\label{NewtonianStrat}
\begin{split}
n_p={}& \dfrac{e^{-(m_p gz+ q \varphi)/T}}{\mathcal{Z}_p} \, , \\
n_e={}& \dfrac{e^{-(m_e gz -q \varphi)/T}}{\mathcal{Z}_e} \, ,\\
\end{split} 
\end{equation}
where $T$ is the temperature, $m_p$ and $m_e$ are the proton and electron masses, $q>0$ is the elementary charge, $\varphi(z)$ is the electric potential, and $\mathcal{Z}_p$, $\mathcal{Z}_e$ are positive constants. Combining \eqref{NewtonianStrat} with the Maxwell equation
%From these expressions, it is evident that local charge neutrality ($n_p = n_e$) cannot be maintained at every $z$ unless the electric potential varies with height. That is, an internal electric field must be present (i.e. $\partial_z \varphi \ne 0$). On the other hand, this electric field must also solve Gauss’s law,
\begin{equation}\label{Coulomblaw}
\partial_z^2 \varphi = -q (n_p {-} n_e) ,
\end{equation}
one obtains a differential equation for $\varphi(z)$, whose solutions generically violate the MHD approximation\footnote{\label{footononon1}The magnitude of the deviations from the MHD limit can be quantified with a quick estimate. The gravitational Boltzmann factor for protons, $e^{-m_pgz/T}$, varies over a length scale 
$L_\text{accel}=T/(m_p g)$. On this same scale, the factor for electrons remains essentially constant, since $e^{-m_egL_\text{accel}/T}=e^{-m_e/m_p}\approx 1$. This mismatch induces significant charge separation when $L_\text{accel}$ becomes smaller than the Debye length $L_\text{Debye}=\sqrt{T/(q^2n_e)}$, the scale over which a plasma fails to maintain charge neutrality \cite[§1.6]{bellan_2006}. This motivates the introduction of a dimensionless ``acceleration number'': $
\mathscr{A} = L_\text{Debye}/L_\text{accel}$. In the solar corona, where $\mathscr{A}\sim 10^{-10}$, MHD is an excellent approximation. However, near a solar-mass black hole, $\mathscr{A}$ may approach unity ($T\sim 10^3$ eV, $n_e\sim 10^9$ cm$^{-3}$, $g=\,$``surface gravity''), thereby falsifying MHD.}, as they can feature both a stratification of charge (protons are heavier than electrons, and tend to accumulate at lower altitudes) and a corresponding electric field, as illustrated by the example in figure \ref{fig:Newtonian}.

The goal of this article is to determine both the electrical polarization and the thermal stratification of \textit{relativistic} plasmas in equilibrium. In particular, we will derive directly from statistical mechanics a system of equations that generalizes \eqref{NewtonianStrat}-\eqref{Coulomblaw}, accounting for all thermal and chemical gradients that are known to develop across thermodynamic equilibria in General Relativity. We will then solve these equations in some simple geometries, and compare the resulting configurations in strong gravity with the Newtonian one shown in figure \ref{fig:Newtonian}.

Throughout the article, we adopt the metric signature $(-,+,+,+)$, and work in ``QFT units'': $c=k_B=\hbar=1$, and $q^2=4\pi/137.036$ (Maxwell's equations in Heaviside-Lorentz units) \cite[\S 54]{Srednicki_2007}. The spacetime may be curved, but it is treated as an externally fixed background, which does not respond to the matter dynamics.

\newpage
\section{Equilibrium states of an accelerating conductor}
\vspace{-0.3cm}

In this section, we derive a (rather natural) variational principle that allows one to determine the equilibrium configurations of an accelerating conductor. The final result agrees with \cite{Hernandez:2017mch,GavassinoShokryMHD:2023qnw}, but the derivation is more detailed, and particular attention is paid to the boundary conditions.

%In this section, we determine from first principles the conditions under which a relativistic conductor undergoing acceleration can attain full thermodynamic equilibrium. Our primary focus is on non-rotational acceleration, such as that of a rocket departing from Earth, or a container suspended near the event horizon of a Schwarzschild black hole.   

\vspace{-0.3cm}
\subsection{Accelerated motions that are compatible with the establishment of equilibrium}\label{IIAAA}
\vspace{-0.3cm}

If we keep shaking a bottle of water chaotically, it will never reach thermodynamic equilibrium. On the other hand, if we leave that bottle on a table (or on the floor of an accelerating rocket), it certainly will. Hence, there is a specific class of accelerating motions that allow for the establishment of a state of thermodynamic equilibrium, and another class that doesn't. Relativistic statistical mechanics gives us an elegant criterion to isolate the former from the latter. Such a criterion is summarized below.

Consider a system of quantum particles confined within a box with perfectly reflecting walls, placed in a background spacetime characterized by the metric tensor $g_{\mu \nu}(x^\alpha)$. Assuming the walls of the box are perfect thermal insulators, their interaction with the quantum particles can be modeled as couplings of the particle's degrees of freedom with a classical, non-dynamical scalar field $\psi(x^\alpha)$, which acts as an externally imposed potential barrier \cite[\S 11]{landau5}. We then say that the set of external conditions $\{g_{\mu \nu}(x^\alpha),\psi(x^\alpha)\}$ is \textit{compatible with the existence of equilibrium} if it is invariant under the action of a group of isometries $x^\mu \rightarrow x^\mu (\tau)$, generated by a timelike Killing vector field $\K^\mu$, namely
\begin{equation}\label{killingone}
\begin{split}
(\mathfrak{L}_\K g)_{\mu \nu}={}& 2\nabla_{(\mu}\K_{\nu)}=0 \, , \\
\mathfrak{L}_\K \psi={}& \K^\mu \nabla_\mu \psi=0 \, ,
\end{split}
\end{equation}
where $\mathfrak{L}$ is the Lie derivative. In fact, if the above holds, we can apply Wigner's theorem \cite[\S 2.2]{weinbergQFT_1995} to the symmetries $x^\mu \rightarrow x^\mu (\tau)$, and conclude that the Hilbert space of the quantum particles must be the carrier space of a family of unitary operators $e^{-iH\tau}$ (with $H\,{=}\, H^\dagger$) which realize the action of such symmetries on all observables, i.e.
\begin{equation}\label{oioioi}
e^{iH\tau}\phi(x^\mu)e^{-iH\tau}=\phi\big(x^\mu (\tau)\big)\, , 
\end{equation}
for any quantum scalar $\phi(x^\mu)$. This allows us to use the generator $H$ to define a microcanonical density matrix,
\begin{equation}\label{grabbino}
\rho_\text{micro} = \dfrac{\Theta(H{-}E)\Theta(E{+}\Delta E{-}H)}{\text{Tr}[\Theta(H{-}E)\Theta(E{+}\Delta E{-}H)]} \spc (\text{for some }E\gg \Delta E>0) \, , 
\end{equation}
which is time-independent, in the sense that $e^{-iH\tau}\rho_\text{micro}e^{iH\tau}=\rho_\text{micro}$, and maximizes the entropy within the space of states with $H\in [E,E{+}\Delta E]$. From this, all the rules of statistical mechanics follow as usual \cite[\S 6.2]{huang_book}, provided that we interpret $H$ as the Hamiltonian and $\rho_\text{micro}$ as the thermodynamic equilibrium state\footnote{It should be noted that, for equation \eqref{grabbino} to be meaningful, the trace of $\Theta(H{-}E)\Theta(E{+}\Delta E{-}H)$ must be finite, which requires $H$ to have some additional properties that make it rather ``energy-like''. For example, we might want the spectrum of $H$ to have a lower bound, otherwise it may be possible to build an infinite number of states with $H=E$ by breaking the interior of the box in two regions, one with $H{\rightarrow} +\infty$ and one with $H{\rightarrow} -\infty$. In appendix \ref{fieldTheory}, we indeed show that, in quantum field theory, the operator $H$ has a standard ``Hamiltonian form'' (provided that $\K^\mu$ is timelike). Thus, $\text{Tr}[\Theta(H{-}E)\Theta(E{+}\Delta E{-}H)]$ is indeed finite, if the box has finite size.}.

In summary: A box moving through a gravitational field can host a medium in thermodynamic equilibrium provided that the geometry of the ``box$+$gravity'' system is invariant under the flow generated by a timelike Killing field $\K^\mu$. This implies, among other things, that the shape of the box should move rigidly across spacetime, since distances and angles are preserved along Killing flows.

\subsection{Non-rotating equilibria}\label{Non-rotazio}

A rocket lifting off from Earth and a spinning bottle of water are both examples of ``accelerating boxes'', namely containers in whose rest frame the enclosed matter is subject to effective gravitational forces. The rotating case, however, is more intricate, as it involves two distinct types of inertial effects: a conventional ``gravity-like'' pulling force (the centrifugal force) and a ``gravitomagnetic'' force (the Coriolis force). To keep the discussion simple, this article will focus on systems such as the departing rocket, where gravitomagnetic effects do not arise. Such non-rotating containers are characterized by the property that their timelike symmetry generator $\K^\mu$ satisfies the condition\footnote{Equation \eqref{Frobenius} is a restatement of Frobenius' hypersurface-orthogonality criterion, $\K_{[\rho} \nabla_{\mu} \K_{\nu]} = 0$ \cite[\S B.3]{Wald}, and its fulfillment for some timelike Killing vector distinguishes static spacetimes (such as Schwarzschild) from stationary but rotating ones (like Kerr) \cite[\S 6.1]{Wald}.}
\begin{equation}\label{Frobenius}
\nabla_{\mu}\K_{\nu}=a_\mu \K_\nu -\K_\mu a_\nu \, , 
\end{equation}
where $a_\mu$ is a spacelike vector field orthogonal to $\K^\mu$ (i.e. $a^\mu \K_\mu = 0$). The equivalence between the validity of \eqref{Frobenius} and the local irrotationality of the container's motion is demonstrated below.

Consider the decomposition $\K_\mu =\K u_\mu$, where $u^\mu$ is a four-velocity (i.e. $u^\mu u_\mu=-1$), and $\K=\sqrt{-\K^\nu \K_\nu}>0$. Then, the second equation of \eqref{killingone} becomes $u^\mu \nabla_\mu \psi=0$, which tells us that the Lagrangian material constituents of the container (in the case of a rocket, such constituents are e.g. screws, rods, wires, panels, and paint) follow worldlines with four-velocity $u^\mu$. Therefore, the kinematic tensors associated with $u^\mu$ reveal the nature of the local transformations experienced by the solid walls of the container. These are easily computed by replacing $\K_\mu$ with $\K u_\mu$ in \eqref{Frobenius}, which gives $\nabla_\mu u_\nu +u_\nu \nabla_\mu \ln \K =a_\mu u_\nu -u_\mu a_\nu$. With a bit of algebra, we obtain
\begin{equation}\label{accelerazione}
\begin{split}
& a_\nu =u^\mu \nabla_\mu u_\nu=\nabla_\nu \ln \K \, , \\
& (g^\alpha_\mu {+}u^\alpha u_\mu)(g^\beta_\nu {+}u^\beta u_\nu)\nabla_\alpha u_\beta = \nabla_\mu u_\nu +u_\mu a_\nu=0 \, . \\
\end{split}
\end{equation}
The first line tells us that $a_\nu(x^\alpha)$ coincides with the proper acceleration experienced by the container at $x^\alpha$, which grows with the spatial gradients of $\ln \K$ (note that $u^\nu \nabla_\nu \K{=}0$). The second line tells us that the container is experiencing (a) no expansion ($\nabla_\mu u^\mu {=}0$), (b) no shear deformation ($\sigma_{\mu \nu}{=}0$), and (c) no vorticity ($\omega_{\mu \nu}{=}0$) \cite[\S 3.1]{rezzolla_book}. Facts (a,b) tell us that the motion of the rocket is rigid (in Born's sense \cite{Born1909}), while (c) tells us that it is irrotational, as claimed.

\subsection{Variational principle}

Now that the external conditions (i.e. the container and the gravitational field) have been fully characterized, we can finally derive the equilibrium condition for a conducting medium placed within such force fields.

Fixed some coarse-graining procedure, consider a discretized list of all possible macroscopic configurations of the conductor of interest, with energy between $E$ and $E{+}\Delta E$, and with a fixed particle numbers. Call such macrostates $\{1,2,...\}$. Each of these macrostates has some microscopic realizations, and we call $\{\mathbb{P}_1,\mathbb{P}_2,...\}$ the corresponding quantum eigenprojectors. Then, the probability for a system described by the density matrix \eqref{grabbino} to be found in the macrostate $n$ is
\begin{equation}
\mathcal{P}_n =\text{Tr}[\mathbb{P}_n\rho_\text{micro}]=\dfrac{\text{Tr}[\mathbb{P}_n]}{\text{Tr}[\Theta(H{-}E)\Theta(E{+}\Delta E{-}H)]} = \dfrac{e^{S_n}}{\sum_l e^{S_l}} \, ,
\end{equation}
where $S_n$ is the entropy of the macrostate $n$ (i.e. the logarithm of its degeneracy). From this, we conclude that the most probable macroscopic configuration of the conductor is the one that maximizes the entropy 
$S$, fixed the values of the ``energy'' $U\,{=}\,\langle H\rangle\, {\approx} \, E$ and the conserved particle numbers, typically the baryon number 
$N_B$ and the electric charge $Q$. In practice, this means that the electrothermal stratification of a conductor described using any coarse-grained theory (e.g. magnetohydrodynamics, or kinetic theory) can be determined via the following variational principle:
\begin{equation}\label{variazionale}
\boxed{\delta S +\alpha^B \delta N_B +\alpha^Q \delta Q +\alpha^U \delta U=0 \, ,}
\end{equation}
where $\{\alpha^B,\alpha^Q,\alpha^U\}$ are some Lagrange multipliers, and ``$\delta$'' is a first-order variation within the state-space of the given coarse-grained theory.

\newpage

\section{Equilibrium conditions for a hydrogen plasma}
\vspace{-0.3cm}

In what follows, we shall apply the variational principle \eqref{variazionale} to a hot, non-degenerate, fully ionized hydrogen plasma, modeled within the framework of relativistic kinetic theory (neglecting spin-polarization effects). This will allow us to rigorously derive the relativistic generalizations of \eqref{NewtonianStrat} and \eqref{Coulomblaw}.

\vspace{-0.3cm}
\subsection{Extensive variables in kinetic theory}
\vspace{-0.3cm}

In kinetic theory \cite{cercignani_book,Groot1980RelativisticKT}, the state of a proton-electron gas is fully characterized by the invariant distribution functions $f_p(x^\alpha,p^\mu)$ and $f_e(x^\alpha,p^\mu)$, which respectively count how many protons and electrons occupy a single-particle quantum state located at the event $x^\alpha$ and with four-momentum $p^\mu$. Then, the extensive variables appearing in \eqref{variazionale} can be expressed as follows:
\begin{equation}\label{SNBQU}
\begin{split}
S ={}& \int_{\Sigma \times \mathcal{H}^3_{m_p}} \left(f_p{-}f_p \ln f_p\right)\, d^6\Gamma +\int_{\Sigma \times \mathcal{H}^3_{m_e}} \left( f_e{-}f_e \ln f_e \right)\, d^6\Gamma \, , \\
N_B ={}& \int_{\Sigma \times \mathcal{H}^3_{m_p}} f_p \, d^6 \Gamma \, , \\
Q ={}& \int_{\Sigma \times \mathcal{H}^3_{m_p}} qf_p\, d^6\Gamma -\int_{\Sigma \times \mathcal{H}^3_{m_e}} qf_e\, d^6\Gamma \, , \\
U ={}& \int_{\Sigma \times \mathcal{H}^3_{m_p}} (\psi-\K^\nu p_\nu) f_p\, d^6\Gamma +\int_{\Sigma \times \mathcal{H}^3_{m_e}} (\psi-\K^\nu p_\nu) f_e\, d^6\Gamma +\int_\Sigma -\K^\nu T^\mu_{\text{em }\nu} d^3\Sigma_\mu \, , \\
\end{split}
\end{equation}
where $\Sigma$ is an arbitrary Cauchy surface (``all of space at one time''), $\mathcal{H}^3_{m_p/m_e}$ is the forward mass hyperboloid of the proton/electron, $d^6\Gamma$ is the density of states in $\Sigma \times \mathcal{H}^3_{m_p/m_e}$, and $d^3\Sigma_\mu$ is the normal surface element of $\Sigma$.

Let us unpack the meaning of \eqref{SNBQU}. The first line is Boltzmann's entropy for non-degenerate gases. In the second line, we expressed the baryon number as the total number of protons, while in the third line we expressed the electric charge as the difference between the total numbers of protons and electrons, multiplied by the elementary charge $q$. In the last line, we used the fact that $U=\langle H\rangle$ is the Noether charge associated with translations generated by $\K^\nu$ in the underlying quantum field theory (see Appendix \ref{fieldTheory}), and thus is the sum of the Noether energies $-\K^\nu p_\nu$ of the particles, plus the Noether energy of the electromagnetic field $F_{\mu \nu}$, whose stress-energy tensor is
\begin{equation}\label{TFF4FF}
T^{\mu }_{\text{em }\nu}=F^{\mu \alpha} F_{\nu \alpha} -\dfrac{1}{4}g\indices{^\mu _\nu} F^{\alpha \beta}F_{\alpha \beta} \, .
\end{equation}
Finally, we added to the energy of both protons and electrons a potential term equal to the external field $\psi$. This allows us to enforce perfect confinement, by setting $\psi = 0$ inside the box, and $\psi = +\infty$ outside.

\vspace{-0.3cm}
\subsection{First-order variations}
\vspace{-0.3cm}

Given that our goal is to solve \eqref{variazionale}, the first step is to compute the variation of \eqref{SNBQU}. Clearly, we have
\begin{equation}\label{dSNBQU}
\begin{split}
\delta S ={}& \int_{\Sigma \times \mathcal{H}^3_{m_p}} {-} \delta f_p \, \ln f_p \, d^6\Gamma +\int_{\Sigma \times \mathcal{H}^3_{m_e}} {-}\delta f_e \ln f_e \, d^6\Gamma \, , \\
\delta N_B ={}& \int_{\Sigma \times \mathcal{H}^3_{m_p}} \delta f_p \, d^6 \Gamma \, , \\
\delta Q ={}& \int_{\Sigma \times \mathcal{H}^3_{m_p}} q\, \delta f_p\, d^6\Gamma -\int_{\Sigma \times \mathcal{H}^3_{m_e}} q\, \delta f_e\, d^6\Gamma \, , \\
\delta U ={}& \int_{\Sigma \times \mathcal{H}^3_{m_p}} (\psi-\K^\nu p_\nu) \delta f_p\, d^6\Gamma +\int_{\Sigma \times \mathcal{H}^3_{m_e}} (\psi-\K^\nu p_\nu)\delta  f_e\, d^6\Gamma +\int_\Sigma -\K^\nu \delta  T^\mu_{\text{em }\nu} d^3\Sigma_\mu \, . \\
\end{split}
\end{equation}
But the perturbation to the energy can be rewritten in a more useful way. In fact, one should keep in mind that the whole variational principle is performed within the space of all \textit{physically admissible} configurations. In particular, both the unperturbed and the perturbed states must obey Maxwell's equations, namely $F_{\mu \nu}=\nabla_\mu A_\nu{-}\nabla_\mu A_\mu$ and $\nabla_\alpha F^{\mu \alpha}=J^\mu$, with $A_\mu$ the four-potential, and $J^\mu$ the electric four-current of the plasma. This fact allows us to rearrange the perturbation to the electromagnetic energy current as follows (see Appendix \ref{apppBBB}):
\newpage
\begin{equation}\label{Temwow}
\begin{split}
\K^\nu \delta T^{\mu }_{\text{em }\nu}={}&  \K^{\nu} A_\nu \delta J^\mu  +2\delta A_\nu \K^{ [\nu}J^{\mu]}+ \delta F^{\mu \alpha} (\mathfrak{L}_{\K} A)_\alpha-\delta A_\alpha (\mathfrak{L}_{\K} F)^{\mu \alpha} \\
&+ \nabla_\alpha(2\K^{[\alpha} F^{\mu] \nu}  \delta A_\nu-\K^{\nu}  \delta A_\nu F^{\mu \alpha}-\K^{\nu} A_\nu \delta F^{\mu \alpha}) \, ,  \\
\end{split}  
\end{equation}
which looks rather complicated, but we can get rid of several terms. First of all, we know from equation \eqref{grabbino} that the equilibrium state is invariant under the spacetime translations generated by $\K^\nu$ (recall: $e^{-iH\tau}\rho_\text{micro}e^{iH\tau}=\rho_\text{micro}$), so that $ \mathfrak{L}_\K \text{``fields''}=0$. This allows us to drop the terms proportional to $\mathfrak{L}_{\K} F$ and $\mathfrak{L}_{\K} A$, in an appropriate gauge. Plugging \eqref{Temwow} into the last line of \eqref{dSNBQU}, and using Stokes' theorem \cite[\S 3.3.3]{PoissonToolkit2009pwt}, we therefore obtain
\begin{equation}\label{laDU}
\begin{split}
\delta U ={}& \int_{\Sigma \times \mathcal{H}^3_{m_p}} \left[\psi-\K^\nu (p_\nu{+}qA_\nu)\right] \delta f_p\, d^6\Gamma +\int_{\Sigma \times \mathcal{H}^3_{m_e}} \left[\psi-\K^\nu (p_\nu{-}qA_\nu)\right]\delta  f_e\, d^6\Gamma \\
-{}& \int_\Sigma  2\delta A_\nu \K^{ [\nu}J^{\mu]} d^3\Sigma_\mu -\dfrac{1}{2}\oint_{\partial \Sigma} (2\K^{[\alpha} F^{\mu] \nu}  \delta A_\nu-\K^{\nu}  \delta A_\nu F^{\mu \alpha}-\K^{\nu} A_\nu \delta F^{\mu \alpha}) d^2 \mathcal{S}_{\mu \alpha} \, .
\end{split}
\end{equation}
Moreover, we can assume that, as $\partial \Sigma$ is pushed to infinity, the boundary integral eventually vanishes. This may be justified, e.g., by assuming that, in a causal gauge (like the Lorenz gauge \cite[\S 10.2]{Wald}), $\delta A_\nu$ has compact support, since it originates from fluctuations of a plasma contained inside a finite box.

\vspace{-0.3cm}
\subsection{Equilibrium distributions}
\vspace{-0.3cm}

Now we are finally ready to determine the conditions for the distribution functions of protons and electrons to be in equilibrium.
Plugging \eqref{dSNBQU} and \eqref{laDU} into \eqref{variazionale}, and recalling that we dropped the integral over $\partial \Sigma$, we obtain
\begin{equation}
\begin{split}
& \int_{\Sigma \times \mathcal{H}^3_{m_p}} ({-}  \ln f_p +\alpha^B+q\alpha^Q +\alpha^U\psi -\alpha^U q\K^\nu A_\nu -\alpha^U \K^\nu p_\nu) \delta f_p \, d^6\Gamma  \\
&+ \int_{\Sigma \times \mathcal{H}^3_{m_e}} ({-}  \ln f_e -q\alpha^Q +\alpha^U\psi +\alpha^U q\K^\nu A_\nu -\alpha^U \K^\nu p_\nu) \delta f_e \, d^6\Gamma  \\
& -2\alpha^U \int_\Sigma  \delta A_\nu \K^{ [\nu}J^{\mu]} d^3\Sigma_\mu =0 \, . \\
\end{split}
\end{equation}
Setting each line to zero, we arrive at the equilibrium conditions we were looking for: 
\begin{equation}\label{fpfe}
    \begin{split}
f_p={}& e^{\alpha^B+q\alpha^Q +\alpha^U\psi -\alpha^U q\K^\nu A_\nu -\alpha^U \K^\nu p_\nu} \, , \\
f_e={}& e^{-q\alpha^Q +\alpha^U\psi +\alpha^U q\K^\nu A_\nu -\alpha^U \K^\nu p_\nu} \, , \\
J^\mu \propto{}& \K^\mu \, . \\
    \end{split}
\end{equation}
Some remarks on the result above are in order before we proceed:
\begin{itemize}
\item It can be easily verified, using the method of the information current \cite{GavassinoGibbs2021}, that the state \eqref{fpfe} is not merely a stationary point of the entropy $S$ (at fixed $N_B$, $Q$, and $U$), but it indeed corresponds to a genuine \textit{maximum}. In other words, we have successfully identified the true state of global thermodynamic equilibrium of the plasma.
\item In order for $f_p$ and $f_e$ to remain finite at high energies, it is necessary that $\alpha^U\,{<}\,0$. As a consequence, the term $\alpha^U\psi$ diverges to $-\infty$ outside the box, ensuring that no particles are present in that region, as expected. Henceforth, unless explicitly stated otherwise, we shall restrict our equations to the interior of the box, where $\psi=0$.
\item The distribution functions \eqref{fpfe} solve the Boltzmann-Vlasov equation \cite[\S 2.1]{cercignani_book}, with vanishing collision integrals. This means that one could have arrived at this same result from dynamical considerations alone. Indeed, \eqref{fpfe} generalizes the equilibrium conditions provided by \cite[Ch. II, \S 4.a, Eq. (24)]{Groot1980RelativisticKT} to accelerating plasmas.
\item Both $f_p$ and $f_e$ are isotropic in the local rest frame identified by the timelike vector $\K^\mu$. As a result, the third line of \eqref{fpfe} becomes redundant. Moreover, this isotropy implies that the collective flow of the plasma must have four-velocity $u^\mu = \K^\mu / \K$, which coincides with the velocity of the walls of the box. It follows that the plasma motion is non-rotating, expansion-free, and shear-free (see Section \ref{Non-rotazio}).
\item The distribution functions \eqref{fpfe} have a Maxwell-J\"{u}ttner (i.e. local equilibrium) form, with local temperature $T\,{=}\,{-}(\alpha^U \K)^{-1}$, and chemical potentials $\mu^p\,{=}\,(\alpha^B{+}q\alpha^Q  {-}\alpha^U q\K^\nu A_\nu)T$ and $\mu^e\,{=}\,({-}q\alpha^Q  {+}\alpha^U q\K^\nu A_\nu)T$. This means that the stratification of the plasma arises from the factor $1/\K$ in the temperature (Tolman law), and from the electromagnetic correction $\propto u^\nu A_\nu$ to the chemical potentials (Lorentz force). These same laws have already been proposed in previous works \cite{Hernandez:2017mch,GavassinoShokryMHD:2023qnw}, but this is the first time they are derived directly from a maximum entropy principle in the context of kinetic theory.
\end{itemize}
 
\subsection{Stratification equations}

We have finally reached the heart of the paper, where the relativistic generalization of \eqref{NewtonianStrat} and \eqref{Coulomblaw} is provided. To keep the notation as close as possible to the Newtonian case, let us introduce the ``redshifted temperature''
$T_\star=-1/\alpha^U$ (which is constant) and the ``electric potential'' $\varphi=-\K^\nu A_\nu$. Then, the distribution functions \eqref{fpfe} become familiar:
\begin{equation}
\begin{split}
f_p ={}& \text{const}\times e^{\frac{\K u^\nu p_\nu-q\varphi}{T_\star}} \, , \\
f_e ={}& \text{const}\times e^{\frac{\K u^\nu p_\nu +q\varphi}{T_\star}} \, . \\
\end{split}    
\end{equation}
Given that these are Maxwell-J\"{u}ttner distributions, the number densities in the rest frame defined by $u^\mu$ can be directly obtained from standard references, such as \cite[Ch.2, \S 4.b, Eq.(33)]{Groot1980RelativisticKT}. For protons, we have
\begin{equation}\label{npfinale}
\boxed{n_p =\dfrac{e^{  {-}\frac{q\varphi}{T_\star}}}{\mathcal{Z}_p \K}  K_2 \left( \dfrac{m_p}{T_\star} \K \right) \, , }
\end{equation}
while for electrons, we have
\begin{equation}\label{nefinale}
\boxed{n_e =\dfrac{e^{+\frac{q\varphi}{T_\star}}}{\mathcal{Z}_e \K}  K_2 \left(\dfrac{m_e}{T_\star} \K \right) \, ,}
\end{equation}
where $K_2$ is the modified Bessel function of the second kind. In the Newtonian limit ($\K \rightarrow 1$ and $m_{p/e}>>T_\star$), we recover \eqref{NewtonianStrat}, with some additional constant prefactors that can be reabsorbed into $\mathcal{Z}_p$ and $\mathcal{Z}_e$.

Deriving a differential equation for $\varphi$ requires a bit more work. First, let us recall that we are working in a gauge where $\mathfrak{L}_\K A=0$, so we can use Cartan's magic formula, and we get $0=\mathfrak{L}_\K A=d(\K \cdot A)+\iota_\K dA=-d\varphi+\iota_\K F$, which in components reads
\begin{equation}\label{nablaphi}
\nabla_\mu \varphi =\K^\nu F_{\nu \mu} \, .
\end{equation}
A more familiar way to write this is $\nabla^\mu \varphi=-\K \mathcal{E}^\mu $, where $\mathcal{E}^\mu$ is the electric field measured by a local observer who comoves with the plasma. Taking the divergence of \eqref{nablaphi}, and invoking \eqref{Frobenius} and \eqref{accelerazione}, we obtain
\begin{equation}
\begin{split}
\nabla^\mu \nabla_\mu \varphi ={}& \nabla^\mu \K^\nu \, F_{\nu \mu} +\K^\nu \nabla^\mu F_{\nu \mu} \\
={}& 2a^\mu \K^\nu F_{\nu \mu} +\K^\nu J_\nu \\
={}& 2 \nabla^\mu {\ln}\, \K \, \, \nabla_\mu \varphi -q\K (n_p-n_e) \, , \\
\end{split}
\end{equation}
or, equivalently,
\begin{equation}\label{finaleGlorioso}
\boxed{ \K^2 \nabla_\mu \left( \K^{-2} \nabla^\mu \varphi\right)  = -q \K (n_p{-}n_e) \, .}
\end{equation}
This is the relativistic generalization of \eqref{Coulomblaw}. Indeed, in the limit where $\K \rightarrow 1$, we recover Poisson's equation\footnote{Note that, even if the differential operator in \eqref{finaleGlorioso} is formally a D'Alembertian, we have that $\K^\mu \nabla_\mu \varphi=0$, which can be proven by contracting \eqref{nablaphi} with $\K^\mu$ (and invoking the antisymmetry of $F_{\mu \nu}$). Hence, there is really no time derivative, and the partial differential equation is effectively Poisson-like.}.

Combining \eqref{npfinale}, \eqref{nefinale}, and \eqref{finaleGlorioso}, one obtains a partial differential equation for $\varphi(x^\alpha)$ which, with appropriate boundary conditions, uniquely determines how charges tend to configure in an accelerating plasma, and the consequent electric field that such charges produce.

In Appendix \ref{appendiceDura}, we also demonstrate that equation \eqref{finaleGlorioso}  provides complete information about all the components of $A_\nu$ in the Lorenz gauge (for which $\nabla_\mu A^\mu \,{=}\, 0$). In fact, we find that
\begin{equation}\label{wowcovaria}
A_\nu = \dfrac{\varphi \K_\nu}{\K^2} \, .
\end{equation}
Moreover, by applying equation \eqref{Frobenius}, one can show that $\K_\nu/\K^2$ is a closed differential form. This allows us, via exterior calculus, to immediately derive the expression for the Faraday tensor:
\begin{equation}\label{FuEEu}
F_{\mu \nu} = u_\mu \mathcal{E}_\nu - \mathcal{E}_\mu u_\nu \, ,
\end{equation}
which confirms that the magnetic field vanishes in the local rest frame of the medium. This is true only because we are dealing with non-rotating systems. 

Note that one can arrive at \eqref{finaleGlorioso} directly by taking the four-divergence of \eqref{FuEEu}, with a bit of algebra.

\section{A couple of applications}
\vspace{-0.3cm}

We now demonstrate how equation \eqref{finaleGlorioso} can be employed to determine the electrothermal stratification of a plasma. To this end, we numerically solve it under suitable boundary conditions in two distinct physical scenarios.

\vspace{-0.3cm}
\subsection{A box of plasma in an accelerating rocket}
\vspace{-0.3cm}

\begin{figure}[h!]
    \centering
    \includegraphics[width=0.50\linewidth]{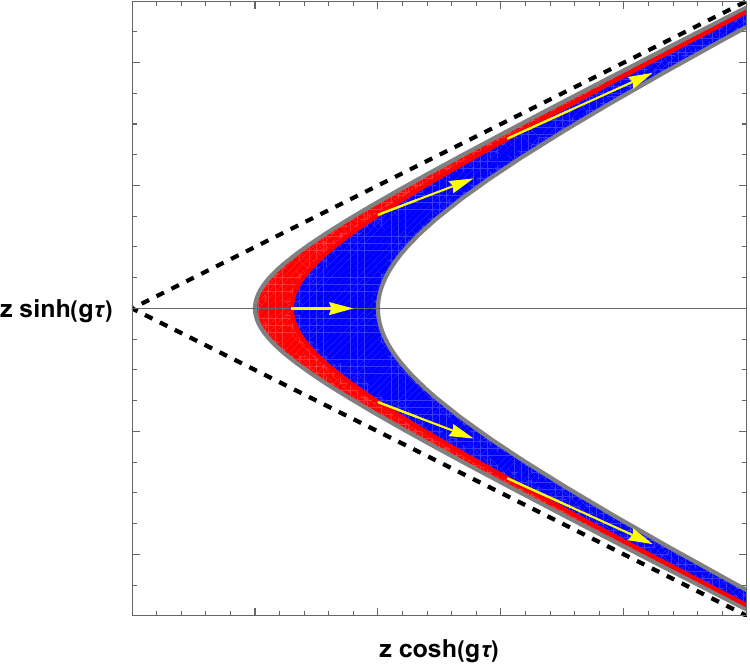}
\caption{Qualitative Minkowski diagram (in Cartesian coordinates) illustrating the equilibrium configuration of a plasma in a container undergoing constant acceleration. The walls of the container (gray) follow hyperbolic motion, which is Born-rigid. Due to their greater mass, protons accumulate towards the rear, leading to a buildup of positive charge (red), while the lighter electrons distribute more evenly throughout the chamber, resulting in an excess of negative charge (blue) towards the front. As a consequence, the plasma is crossed by an electric field (yellow arrows) in its local rest frame. The dashed lines mark the boundary of the Rindler wedge, outside of which the curvilinear coordinates $\tau$ and $z$ lose any meaning.}
    \label{fig:Qualitative}
\end{figure}

We examine the physical setting shown in figure \ref{fig:Qualitative}: A plasma layer is confined between two infinitely extended plates undergoing hyperbolic motion in Minkowski space. 
As is typical for problems involving constant acceleration, the most convenient coordinate system is Rindler's variables, which result in the metric being
$ds^2=-g^2 z^2 d\tau^2 +dx^2+dy^2+dz^2$. In these coordinates, the plates are positioned at some stationary locations $z=z_\pm$, with $0<z_-<z_+$. The timelike symmetry generator of the problem is $\K^\mu \partial_\mu=\partial_\tau$ (i.e. the generator of boosts), so that $u^\mu=(gz)^{-1}\delta^\mu_\tau$ and $\K=gz$. Consequently, the proper acceleration felt by the volume elements of plasma (which move with velocity $u^\mu$) is $a=1/z$. This allows us to interpret $g$ as the acceleration $a$ felt by the volume element whose proper time equals $\tau$.

In this geometry, equation \eqref{finaleGlorioso} reduces to
\begin{equation}\label{ventisei}
z\, \partial_z \left( z^{-1} \partial_z \varphi\right)  = -q \left[\dfrac{e^{  {-}\frac{q\varphi}{T_\star}}}{\mathcal{Z}_p }  K_2 \left( \dfrac{m_p gz}{T_\star} \right){-}\dfrac{e^{+\frac{q\varphi}{T_\star}}}{\mathcal{Z}_e }  K_2 \left(\dfrac{m_e gz}{T_\star} \right)\right] \, ,
\end{equation}
where we have used the fact that $\partial_\tau \varphi=\K^\mu \partial_\mu \varphi=0$. As boundary conditions, we assume that the system is globally neutral, and there is no externally imposed electric field, so that $\varphi'(z_-)=\varphi'(z_+)=0$. In order to solve \eqref{ventisei}, it is convenient to introduce the dimensionless quantities $\Tilde{\varphi}=q\varphi/T_\star$, $\Tilde{z}=m_p g z/T_\star$, and $\Tilde{\mathcal{Z}}_{p/e}= m_p^2 g^2 \mathcal{Z}_{p/e}/(q^2 T_\star)$, giving 
\begin{equation}\label{rescadina}
\Tilde{z} \, \partial_{\Tilde{z}} \left( \Tilde{z}^{-1} \, \partial_{\Tilde{z}} \Tilde{\varphi}\right)  = -\dfrac{e^{  {-}\Tilde{\varphi}}}{\Tilde{\mathcal{Z}}_p }  K_2 \left( \Tilde{z} \right){+}\dfrac{e^{\Tilde{\varphi}}}{\Tilde{\mathcal{Z}}_e }  K_2 \left(\dfrac{m_e}{m_p}\Tilde{z}  \right) \, .
\end{equation}
Notably, the coordinate $\Tilde{z}$ equals the local ``coldness'' of the proton gas (i.e. $m_p/T$), so that lower values of $\Tilde{z}$ indicate higher temperatures. In particular, $\Tilde{z} \sim 1$ identifies the threshold at which protons transition to relativistic behavior\footnote{Strictly speaking, equation \eqref{finaleGlorioso} ceases to be valid when $T \gtrsim m_{e}$, as pair-production processes begin to alter the right-hand side \cite{landau5}. However, we disregard this effect, since it would distract us from the purpose of the work.}. 

\begin{figure}[h!]
    \centering
\includegraphics[width=0.49\linewidth]{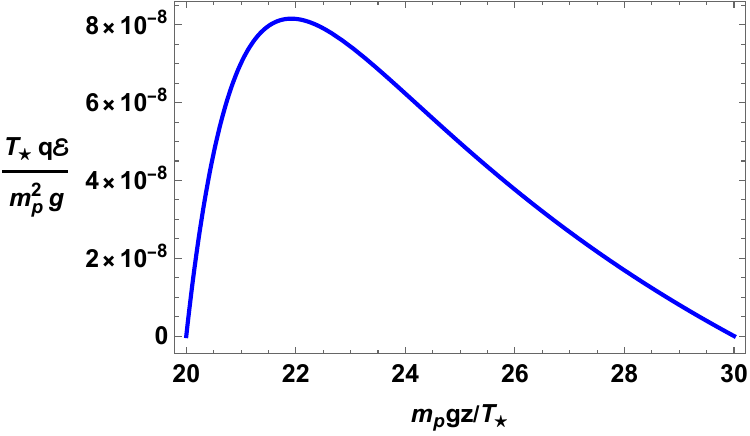}
\includegraphics[width=0.49\linewidth]{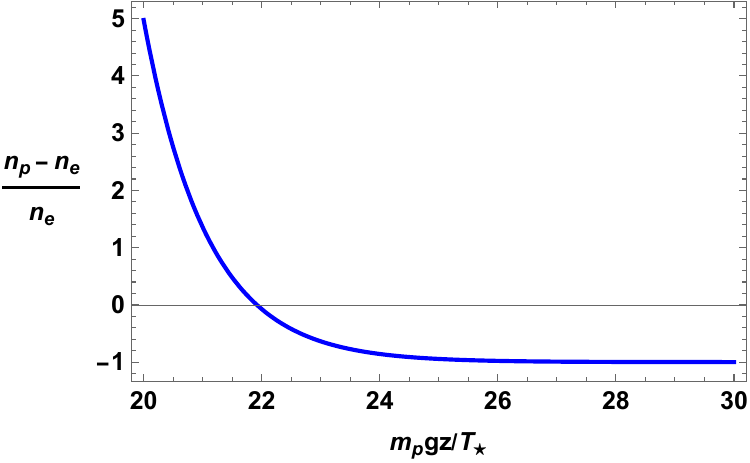}
\includegraphics[width=0.49\linewidth]{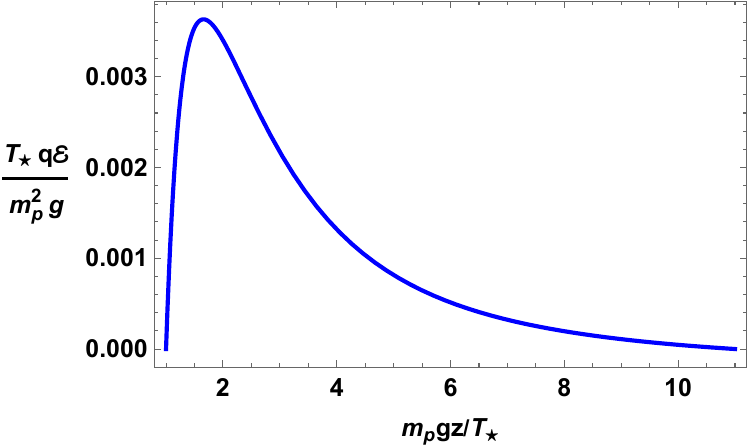}
\includegraphics[width=0.49\linewidth]{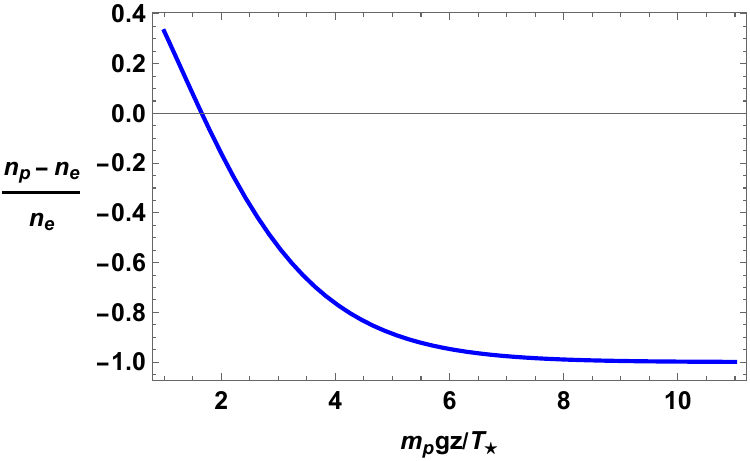}
\includegraphics[width=0.49\linewidth]{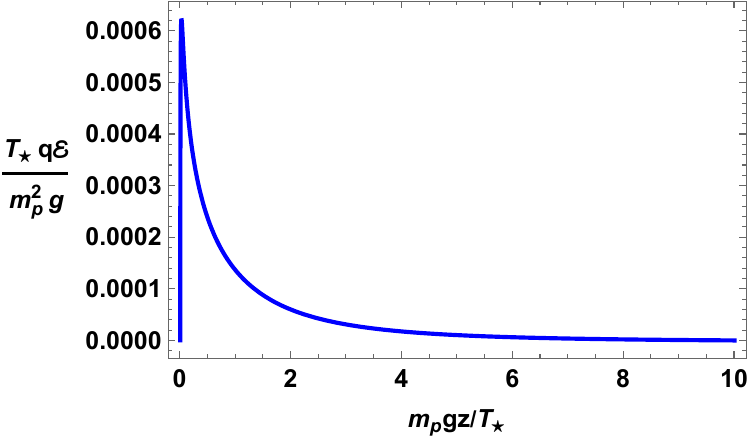}
\includegraphics[width=0.49\linewidth]{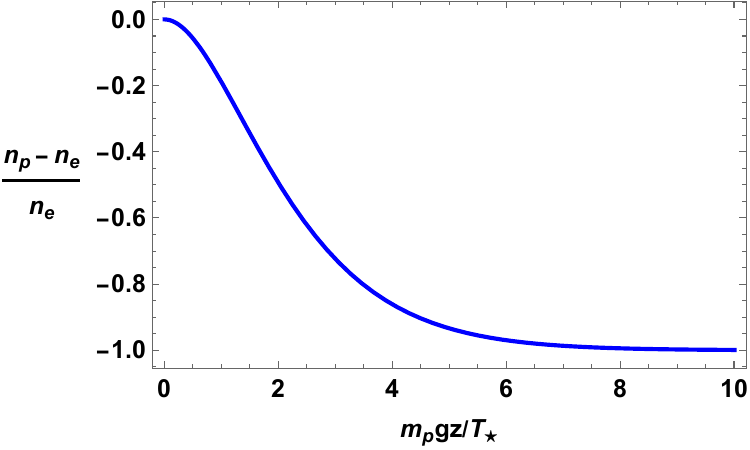}
\caption{Equilibrium electric field (left panels) and charge separation (right panels) of a plasma column undergoing hyperbolic motion as in figure \ref{fig:Qualitative}. Each row corresponds to a different solution of \eqref{rescadina}, with $\{\Tilde{z}_-,\Tilde{z}_+\}$ equal to $\{20,30\}$ (upper panels), $\{1,11\}$ (middle panels), and $\{0.01,10.01\}$ (lower panels). All solutions assume that $\Tilde{\mathcal{Z}}_p=20$ and $\Tilde{\mathcal{Z}}_e=20 \, K_2(\frac{m_e}{m_p}\Tilde{z}_-)$.}
    \label{fig:RockettiamoPlots}
\end{figure}

In Figure \ref{fig:RockettiamoPlots}, we display several numerical solutions of equation \eqref{rescadina}, each corresponding to a progressively more relativistic regime. This is achieved by shifting an interval of fixed width $\Delta \Tilde{z} = 10$ closer to the Rindler horizon (namely, by taking $\Tilde{z}_- \rightarrow 0^+$). As a result, the rear of the rocket undergoes increasingly intense proper acceleration, and the plasma heats up significantly. Notably, the species most affected by relativistic thermodynamics are the electrons. In the Newtonian regime (figure \ref{fig:Newtonian}), the proton density decreases rapidly under the influence of gravity (or inertia), while the electrons remain uniformly distributed over the same lengthscale. In the ultrarelativistic regime, however, since $n_e \propto T^3$, the strong thermal gradients lead to a $z^{-3}$ modulation of the electron density, causing the electrons to accumulate at the rear as well. Consequently, the ratio $(n_p-n_e)/n_e$ tends to $0^+$ at $z_-$ in the ultrarelativistic limit: the difference $n_p {-} n_e$ stays finite (and positive), while both densities diverge in the same manner.

% Now, we recall that our main goal is to study how the temperature gradients (a purely relativistic phenomenon) modify the electric field in the plasma. This motivates us to introduce the dimensionless number
% \begin{equation}
% \mathcal{R}=\dfrac{T(z_-)-T(z_+)}{T(z_+)}=\dfrac{z_+-z_-}{z_-} \, ,
% \end{equation}
% which quantifies the overall importance of relativistic thermodynamic effects in the plasma. In figure \ref{fig:RockettiamoPlots}, we consider scenarios with increasing values of $\mathcal{R}$. As expected, when $\mathcal{R}\lesssim 1$, the profile is similar to that of figure \ref{fig:Newtonian}. As $\mathcal{R}\gg 1$, two things happen. First of all, the proper acceleration at $z_-$ becomes much higher than that at $z_+$, and this causes both positive and negative charges to accumulate towards the rear.

\subsection{A spherical shell of plasma suspended above a black hole horizon}
\vspace{-0.3cm}

\begin{figure}[h!]
    \centering
\includegraphics[width=0.49\linewidth]{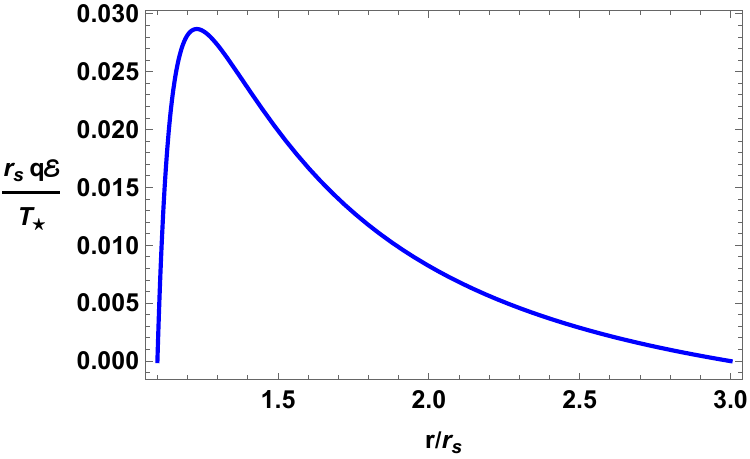}
\includegraphics[width=0.49\linewidth]{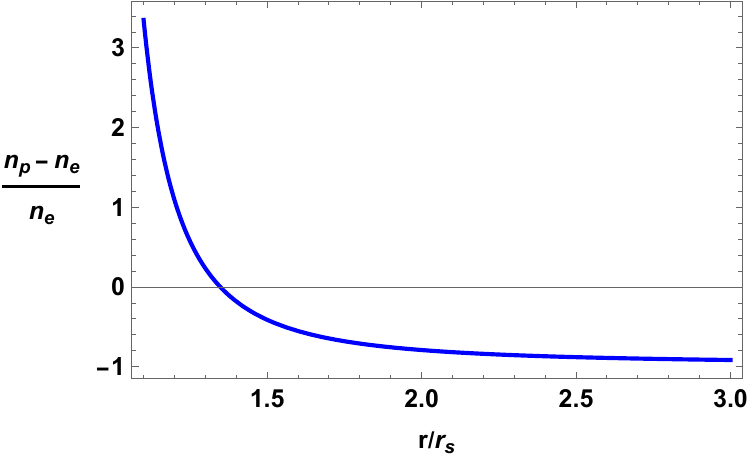}
\includegraphics[width=0.49\linewidth]{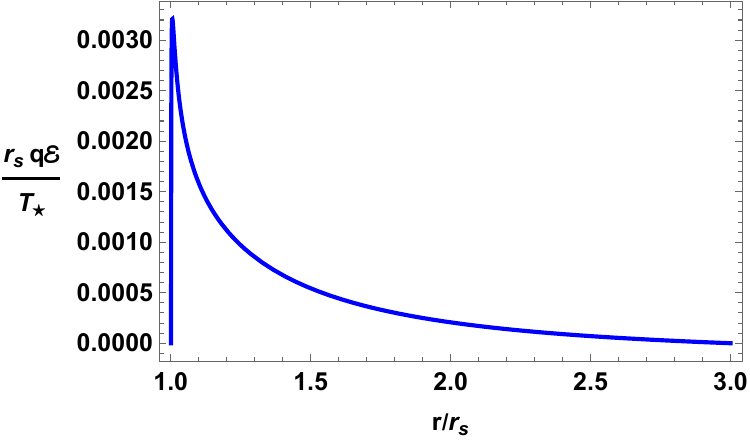}
\includegraphics[width=0.49\linewidth]{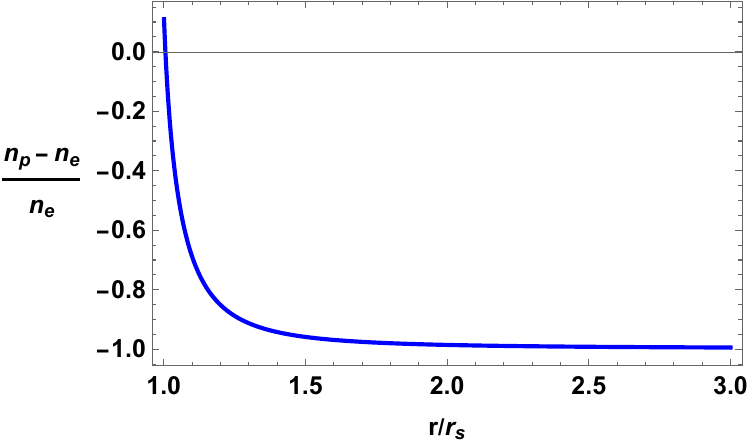}
\caption{Equilibrium electric field (left panels) and charge separation (right panels) of a plasma shell suspended above the event horizon of a Schwarzschild black hole. The two rows corresponds to different solutions of \eqref{eugualeaprima}, with $\{\Tilde{r}_-,\Tilde{r}_+\}$ equal to $\{1.1,3\}$ (upper panels), and $\{1.001,3\}$ (lower panels). Both solutions assume that $m_p=10 T_\star$ and $\mathcal{Z}_{p/e}=20 \frac{r_s^2 q^2}{T_\star} K_2\big(\frac{m_{p/e}}{T_\star}\sqrt{1{-}\frac{r_s}{r_-}}\big)$.}
    \label{fig:placebo}
\end{figure}

In our second example, we consider a non-rotating black hole with Schwarzschild radius $r_s$, described by the metric
\begin{equation}
ds^2 =-\left(1-\dfrac{r_s}{r}\right) d\tau^2 + \left(1-\dfrac{r_s}{r}\right)^{-1} dr^2 +r^2(d\theta^2 +\sin^2\theta \, d\phi^2)  \, ,
\end{equation}
which is surrounded by a plasma enclosed between two concentric spherical shells of solid matter located at $r=r_\pm$, where $r_s<r_-<r_+$. Assuming that the plasma is at rest, the relevant symmetry generator of the problem is the Killing vector 
$\K^\mu \partial_\mu =\partial_\tau$, so that $u^\mu=\delta^\mu_\tau /\sqrt{1{-}r_s/r}$, and $\K=\sqrt{1{-}r_s/r \,}$. In this geometry, equation \eqref{finaleGlorioso} becomes
\begin{equation}\label{eugualeaprima}
\dfrac{1}{r^2} \partial_r \left( r^2 \partial_r \varphi\right)  = -q\left(1{-}\dfrac{r_s}{r}\right)^{-1}  \left[\dfrac{e^{  {-}\frac{q\varphi}{T_\star}}}{\mathcal{Z}_p }  K_2 \left( \dfrac{m_p}{T_\star} \sqrt{1{-}\dfrac{r_s}{r}} \right){-}\dfrac{e^{+\frac{q\varphi}{T_\star}}}{\mathcal{Z}_e }  K_2 \left(\dfrac{m_e}{T_\star} \sqrt{1{-}\dfrac{r_s}{r}} \right)\right] \, ,
\end{equation}
which should be solved with boundary conditions $\varphi'(r_-)=\varphi'(r_+)=0$ (again, assuming global charge neutrality). In figure \eqref{fig:placebo}, we provide a couple of examples. As one may have expected, equation \eqref{eugualeaprima} exhibits behavior closely resembling that of \eqref{ventisei}, with the black hole's event horizon effectively taking the place of the Rindler horizon. Also in this case, the electric field is dramatically reduced as $r_-$ approaches the event horizon, because the electrons are pulled towards the black hole very strongly, and they nearly neutralize the  plasma.

% Once again, assuming that both the compact object and the plasma are globally neutral, the boundary conditions are given by $r^2\partial_r\varphi=0$ at both $r_-$ and infinity.
% \begin{equation}
% \dfrac{1}{r^2} \partial_r \left( r^2 \partial_r \varphi\right)  = -q  \left[\sqrt{\dfrac{\pi T_\star}{2 m_p}}\dfrac{e^{- \frac{m_p}{T_\star}}}{\mathcal{Z}_p } e^{  {-}\frac{q\varphi}{T_\star}+ \frac{m_p M}{T_\star r} } {-} \sqrt{\dfrac{\pi T_\star}{2 m_e}} \dfrac{e^{-\frac{m_e}{T_\star}}}{\mathcal{Z}_e } e^{\frac{q\varphi}{T_\star} +\frac{m_e M}{T_\star r} } \right] \, .
% \end{equation}

\vspace{-0.2cm}
\section{Conclusions}
\vspace{-0.2cm}

In this work, we have shown that accelerating relativistic plasmas naturally develop equilibrium configurations featuring both charge separation and rest-frame electric fields, in direct violation of the assumptions of both ideal and resistive MHD. We derived a general stratification equation [eq. \eqref{finaleGlorioso}] governing this behavior, based solely on the maximum entropy principle in relativistic kinetic theory. Our results generalize the classical Stewart-Tolman effects to the fully relativistic domain and curved spacetime geometries, accounting also for equilibrium temperature gradients.

We demonstrated how relativistic acceleration modifies both the charge and thermal structure of a plasma by solving the stratification equation in two distinct settings: a uniformly accelerated box in Minkowski space, and a spherical shell suspended above a Schwarzschild black hole. In both cases, we found that the electric field is non-vanishing in equilibrium, and its value depends on the plasma's equation of state and geometry, but not on its transport coefficients. 
\textit{These electric fields survive even in a plasma with infinite conductivity}.

Our findings highlight the limitations of standard MHD in extreme environments and provide a robust theoretical framework for modeling plasmas near black holes, in relativistic jets, and possibly in laboratory setups where strong accelerations are achieved. Future work may include exploring time-dependent generalizations, the role of pair production at higher temperatures, and quantitative comparisons with charge-separating MHD models (like \cite{GavassinoShokryMHD:2023qnw}) in dynamical scenarios.

% When a plasma experiences extremely high accelerations (see footnote \ref{footononon1} for the meaning of ``extremely high''), standard MHD breaks down, as rest-frame charge densities and electric fields emerge. As shown in this article, these fields persist even in thermodynamic equilibrium, and their values are determined solely by the plasma's equation of state, not by its electrical conductivity. In practice, such large accelerations are observed only in low-density plasmas under strong gravity, or in high-energy plasmas undergoing violent transformations.  

% \newpage
% Relativistic magnetohydrodynamic frameworks that permit spontaneous charge separation have already been developed and applied to the quark-gluon plasma. Building on the analysis in \cite{GavassinoShokryMHD:2023qnw}, one can readily verify that the maximum-entropy configuration in these theories is consistent with equations \eqref{npfinale}, \eqref{nefinale}, and \eqref{finaleGlorioso}. It would be interesting to investigate numerically the extent to which such theories remain more accurate than standard MHD in a regime of transient acceleration.

\vspace{-0.2cm}
\section*{Acknowledgements}
\vspace{-0.2cm}

This work is partially supported by a Vanderbilt Seeding Success Grant.

\appendix

\vspace{-0.2cm}
\section{Derivation of the quantum Hamiltonian from field theory}\label{fieldTheory}
\vspace{-0.2cm}

Let $\phi_i(x^\alpha)$ be a list of quantum fields that we use to characterize the microphysics dynamics of the system, and recall that the external conditions are described as a coupling with the classical scalar field $\psi(x^\alpha)$, which acts as an externally-fixed ``source term'' \cite{GavassinoCovarianceThermo:2021kia}. In field theory, the microscopic action of the system takes the general form
\begin{equation}\label{ActionandAction}
\mathcal{I}=\int \mathcal{L}(\phi_i,\partial_\mu \phi_i,g_{\mu \nu},\psi)\sqrt{-g}\, d^4 x \, ,
\end{equation}
where $\mathcal{L}$ is a Lagrangian density. As usual, the Euler-Lagrange equations read
\begin{equation}\label{EulerLagrange}
\sqrt{-g} \,\dfrac{\partial \mathcal{L}}{\partial \phi_i}-\partial_\mu \left(\sqrt{-g} \,\dfrac{\partial \mathcal{L}}{\partial (\partial_\mu \phi_i)} \right) =0\, .
\end{equation}
Now, let us work in a system of coordinates $\{\tau,x^1,x^2,x^3\}$ where $\K^\mu \partial_\mu =\partial_\tau$. Then, equation \eqref{killingone} becomes
\begin{equation}\label{killingone2}
\begin{split}
\partial_\tau g_{\mu \nu}={}& 0 \, , \\
\partial_\tau \psi={}& 0 \, .
\end{split}
\end{equation}
Hence, $\mathcal{L}$  does not depend explicitly on time. This allows us to define the following Noether charge:
\begin{equation}\label{HHHamilt}
H =\int \left[\dfrac{\partial \mathcal{L}}{\partial(\partial_\tau \phi_i)} \partial_\tau \phi_i -\mathcal{L} \right] \sqrt{-g}\, dx^1 dx^2 dx^3 \, ,
\end{equation}
which is conserved (i.e. $dH/d\tau=0$) thanks to \eqref{EulerLagrange}, and depends on $\tau$ only through $\phi_i$. This is precisely the Hamiltonian of the system relative to the time coordinate $\tau$, since the action \eqref{ActionandAction} can be expressed as the integral of a Lagrangian, $\mathcal{I}=\int L d\tau$, with
\begin{equation}
L=\int \mathcal{L}\sqrt{-g}\, dx^1 dx^2 dx^3 \, ,
\end{equation}
which shows that $L$ and $H$ are Legendre transforms of each other, relative to the variables $\partial_\tau \phi_i$.

At this point, it only remains to prove that the Hamiltonian \eqref{HHHamilt} coincides with the operator $H$ introduced in section \ref{IIAAA}. To this end, we differentiate \eqref{oioioi} with respect to $\tau$. This gives $\partial_\tau \phi =i\left[H,\phi\right]$, which is indeed obeyed by \eqref{HHHamilt}, assuming canonical commutation relations.

\vspace{-0.2cm}
\section{Variation of the electromagnetic energy current}\label{apppBBB}
\vspace{-0.2cm}

We start by contracting \eqref{TFF4FF} with $\K^\nu$, and we obtain
\begin{equation}
 \K^\nu T^{\mu }_{\text{em }\nu}= \K^{\nu} F^{\mu \alpha} F_{\nu \alpha} -\dfrac{1}{4} \K^{\mu} F^{\alpha \beta}F_{\alpha \beta} \, ,   
\end{equation}
whose first-order variation is
\begin{equation}\label{KTtt}
\K^\nu \delta  T^{\mu }_{\text{em }\nu}= \K^{\nu} \delta F^{\mu \alpha} F_{\nu \alpha}+ \K^{\nu} F^{\mu \alpha} \delta F_{\nu \alpha} -\dfrac{1}{2} \K^{\mu} F^{\alpha \beta}\delta F_{\alpha \beta} \, .
\end{equation}
We then assume that Maxwell's equations hold, both for the unperturbed state
($F_{\mu \nu}\,{=}\,\nabla_\mu A_\nu{-}\nabla_\nu A_\mu$ and  $\nabla_\alpha F^{\mu \alpha}\,{=}\,J^\mu$) and for the perturbation ($\delta F_{\mu \nu}\,{=}\,\nabla_\mu \delta A_\nu{-}\nabla_\nu \delta A_\mu$ and  $\nabla_\alpha \delta F^{\mu \alpha}\,{=}\,\delta J^\mu$). Hence, the first piece of \eqref{KTtt} becomes
\begin{equation}
\begin{split}
\K^{\nu} \delta F^{\mu \alpha} F_{\nu \alpha} ={}& \K^{\nu} \delta F^{\mu \alpha} \nabla_\nu A_\alpha - \K^{\nu} \delta F^{\mu \alpha}\nabla_\alpha A_\nu \\
={}&  \delta F^{\mu \alpha} \K^{\nu} \nabla_\nu A_\alpha - \nabla_\alpha(\K^{\nu} \delta F^{\mu \alpha} A_\nu)+\delta F^{\mu \alpha} A_\nu \nabla_\alpha \K^{\nu}+\K^{ \nu}A_\nu \nabla_\alpha \delta F^{\mu \alpha} \\ 
={}&  \delta F^{\mu \alpha} (\mathfrak{L}_{\K} A)_\alpha - \nabla_\alpha(\K^{\nu} A_\nu \delta F^{\mu \alpha} )+ \K^{ \nu}A_\nu \delta J^\mu \, , \\ 
\end{split}
\end{equation}
while the second becomes
\begin{equation}
\begin{split}
\K^{\nu} F^{\mu \alpha} \delta F_{\nu \alpha} ={}& \K^{\nu} F^{\mu \alpha} \nabla_\nu \delta A_\alpha -\K^{\nu} F^{\mu \alpha} \nabla_\alpha \delta A_\nu  \\
={}& \nabla_\nu(\K^{\nu} F^{\mu \alpha}  \delta A_\alpha)-\delta A_\alpha \K^{\nu}\nabla_\nu F^{\mu \alpha} - \nabla_\alpha(\K^{\nu} F^{\mu \alpha}  \delta A_\nu)+\delta A_\nu F^{\mu \alpha}\nabla_\alpha \K^{ \nu} +\delta A_\nu \K^{ \nu}\nabla_\alpha F^{\mu \alpha}  \\
={}& \nabla_\alpha(\K^{\alpha} F^{\mu \nu}  \delta A_\nu)-\delta A_\alpha \K^{\nu}\nabla_\nu F^{\mu \alpha} - \nabla_\alpha(\K^{\nu} F^{\mu \alpha}  \delta A_\nu)+\delta A_\alpha F^{\mu \nu}\nabla_\nu \K^{ \alpha} +\delta A_\nu \K^{ \nu}\nabla_\alpha F^{\mu \alpha}  \\
={}& \nabla_\alpha(\K^{\alpha} F^{\mu \nu}  \delta A_\nu-\K^{\nu} F^{\mu \alpha}  \delta A_\nu)-\delta A_\alpha (\K^{\nu}\nabla_\nu F^{\mu \alpha}- F^{\mu \nu}\nabla_\nu \K^{ \alpha}) +\delta A_\nu \K^{ \nu}J^\mu \, ,  \\
\end{split}
\end{equation}
and the third becomes
\begin{equation}
\begin{split}
-\dfrac{1}{2} \K^{\mu} F^{\alpha \beta}\delta F_{\alpha \beta} ={}& -\dfrac{1}{2} \K^{\mu} F^{\alpha \beta}(\nabla_\alpha \delta A_\beta -\nabla_\beta \delta A_\alpha) \\
={}&  \K^{\mu} F^{\alpha \nu} \nabla_\nu \delta A_\alpha \\
={}& \nabla_\nu( \K^{\mu} F^{\alpha \nu}  \delta A_\alpha)-\delta A_\alpha F^{\alpha \nu}\nabla_\nu \K^{ \mu} -\K^{\mu}  \delta A_\alpha \nabla_\nu F^{\alpha \nu} \\
={}& -\nabla_\alpha( \K^{\mu} F^{\alpha \nu}  \delta A_\nu)+\delta A_\alpha F^{\nu\alpha}\nabla_\nu \K^{ \mu} -\K^{\mu}  \delta A_\alpha J^\alpha \, .\\
\end{split}
\end{equation}
Combining everything together, we obtain
\begin{equation}
\begin{split}
\K^\nu \delta  T^{\mu }_{\text{em }\nu}={}&  \K^{ \nu}A_\nu \delta J^\mu  +2\delta A_\nu \K^{ [\nu}J^{\mu]}+ \delta F^{\mu \alpha} (\mathfrak{L}_{\K} A)_\alpha-\delta A_\alpha (\mathfrak{L}_{\K} F)^{\mu \alpha} \\
&+ \nabla_\alpha(2\K^{[\alpha} F^{\mu] \nu}  \delta A_\nu-\K^{\nu}  \delta A_\nu F^{\mu \alpha}-\K^{\nu} A_\nu \delta F^{\mu \alpha})  \, . \\
\end{split}  
\end{equation}

\section{Maxwell's equations in adapted coordinates}\label{appendiceDura}

It is well known \cite[\S 6.1]{Wald} that, if a Killing vector obeys \eqref{Frobenius}, then there is a system of coordinates such that
\begin{equation}
\begin{split}
ds^2 ={}& -\K^2(x^1,x^2,x^3) \, d\tau^2 +g_{jk}(x^1,x^2,x^3) \, dx^j \, dx^k \, , \\
\K^\mu \partial_\mu ={}& \partial_\tau \, .\\
\end{split}
\end{equation}
Let us just \textit{postulate} that $A\,{=}\,{-}\varphi \, d\tau$, which is consistent with the definition of $\varphi$, and becomes \eqref{wowcovaria} in generic coordinates. Its exterior derivative gives
\begin{equation}
F = -\partial_j \varphi \, dx^j \wedge d\tau \, .
\end{equation}
Let us now write down the four-divergence
$\nabla_\mu F^{\mu \nu}$ in the coordinates above:
\begin{equation}
\dfrac{1}{\sqrt{-g}} \partial_\mu (\sqrt{-g} \, g^{\mu \lambda} g^{\nu \sigma}F_{\lambda \sigma})=-\dfrac{1}{\sqrt{-g}} \partial_j (\sqrt{-g} \, g^{j k} g^{\nu \tau}\partial_k \varphi)=\dfrac{\delta^\nu_\tau}{\sqrt{-g}} \partial_j (\sqrt{-g} \, g^{j k} \K^{-2}\partial_k \varphi) \, .
\end{equation}
We see that $\nabla_\mu F^{\mu \nu}\propto \K^\nu$, which shows that our postulate for $A$ is consistent with the third line of \eqref{fpfe}. The only relevant equation that remains is $\nabla_\mu F^{\mu \tau}=-q(n_p-n_e)u^\tau$ which, multiplied by $\K^2$, gives
\begin{equation}
\dfrac{\K^2}{\sqrt{-g}} \partial_j (\sqrt{-g} \, g^{j k} \K^{-2} \partial_k \varphi)= -q\K (n_p{-}n_e) \, . 
\end{equation}
This equation is \eqref{finaleGlorioso} expressed in the above coordinates.

\bibliography{Biblio}

%merlin.mbs apsrev4-1.bst 2010-07-25 4.21a (PWD, AO, DPC) hacked
%Control: key (0)
%Control: author (72) initials jnrlst
%Control: editor formatted (1) identically to author
%Control: production of article title (-1) disabled
%Control: page (0) single
%Control: year (1) truncated
%Control: production of eprint (0) enabled
\begin{thebibliography}{32}%
\makeatletter
\providecommand \@ifxundefined [1]{%
 \@ifx{#1\undefined}
}%
\providecommand \@ifnum [1]{%
 \ifnum #1\expandafter \@firstoftwo
 \else \expandafter \@secondoftwo
 \fi
}%
\providecommand \@ifx [1]{%
 \ifx #1\expandafter \@firstoftwo
 \else \expandafter \@secondoftwo
 \fi
}%
\providecommand \natexlab [1]{#1}%
\providecommand \enquote  [1]{``#1''}%
\providecommand \bibnamefont  [1]{#1}%
\providecommand \bibfnamefont [1]{#1}%
\providecommand \citenamefont [1]{#1}%
\providecommand \href@noop [0]{\@secondoftwo}%
\providecommand \href [0]{\begingroup \@sanitize@url \@href}%
\providecommand \@href[1]{\@@startlink{#1}\@@href}%
\providecommand \@@href[1]{\endgroup#1\@@endlink}%
\providecommand \@sanitize@url [0]{\catcode `\\12\catcode `\$12\catcode `\&12\catcode `\#12\catcode `\^12\catcode `\_12\catcode `\%12\relax}%
\providecommand \@@startlink[1]{}%
\providecommand \@@endlink[0]{}%
\providecommand \url  [0]{\begingroup\@sanitize@url \@url }%
\providecommand \@url [1]{\endgroup\@href {#1}{\urlprefix }}%
\providecommand \urlprefix  [0]{URL }%
\providecommand \Eprint [0]{\href }%
\providecommand \doibase [0]{http://dx.doi.org/}%
\providecommand \selectlanguage [0]{\@gobble}%
\providecommand \bibinfo  [0]{\@secondoftwo}%
\providecommand \bibfield  [0]{\@secondoftwo}%
\providecommand \translation [1]{[#1]}%
\providecommand \BibitemOpen [0]{}%
\providecommand \bibitemStop [0]{}%
\providecommand \bibitemNoStop [0]{.\EOS\space}%
\providecommand \EOS [0]{\spacefactor3000\relax}%
\providecommand \BibitemShut  [1]{\csname bibitem#1\endcsname}%
\let\auto@bib@innerbib\@empty
%</preamble>
\bibitem [{\citenamefont {Banerjee}\ \emph {et~al.}(2012)\citenamefont {Banerjee}, \citenamefont {Bhattacharya}, \citenamefont {Bhattacharyya}, \citenamefont {Jain}, \citenamefont {Minwalla},\ and\ \citenamefont {Sharma}}]{Banerjee:2012iz}%
  \BibitemOpen
  \bibfield  {author} {\bibinfo {author} {\bibfnamefont {N.}~\bibnamefont {Banerjee}}, \bibinfo {author} {\bibfnamefont {J.}~\bibnamefont {Bhattacharya}}, \bibinfo {author} {\bibfnamefont {S.}~\bibnamefont {Bhattacharyya}}, \bibinfo {author} {\bibfnamefont {S.}~\bibnamefont {Jain}}, \bibinfo {author} {\bibfnamefont {S.}~\bibnamefont {Minwalla}}, \ and\ \bibinfo {author} {\bibfnamefont {T.}~\bibnamefont {Sharma}},\ }\href {\doibase 10.1007/JHEP09(2012)046} {\bibfield  {journal} {\bibinfo  {journal} {JHEP}\ }\textbf {\bibinfo {volume} {09}},\ \bibinfo {pages} {046} (\bibinfo {year} {2012})},\ \Eprint {http://arxiv.org/abs/1203.3544} {arXiv:1203.3544 [hep-th]} \BibitemShut {NoStop}%
\bibitem [{\citenamefont {Becattini}\ and\ \citenamefont {Grossi}(2015)}]{BecattiniQuantumCorrections2015}%
  \BibitemOpen
  \bibfield  {author} {\bibinfo {author} {\bibfnamefont {F.}~\bibnamefont {Becattini}}\ and\ \bibinfo {author} {\bibfnamefont {E.}~\bibnamefont {Grossi}},\ }\href {\doibase 10.1103/PhysRevD.92.045037} {\bibfield  {journal} {\bibinfo  {journal} {Phys. Rev. D}\ }\textbf {\bibinfo {volume} {92}},\ \bibinfo {pages} {045037} (\bibinfo {year} {2015})}\BibitemShut {NoStop}%
\bibitem [{\citenamefont {{Becattini}}(2016)}]{BecattiniBeta2016}%
  \BibitemOpen
  \bibfield  {author} {\bibinfo {author} {\bibfnamefont {F.}~\bibnamefont {{Becattini}}},\ }\href {\doibase 10.5506/APhysPolB.47.1819} {\bibfield  {journal} {\bibinfo  {journal} {Acta Physica Polonica B}\ }\textbf {\bibinfo {volume} {47}},\ \bibinfo {pages} {1819} (\bibinfo {year} {2016})},\ \Eprint {http://arxiv.org/abs/1606.06605} {arXiv:1606.06605 [gr-qc]} \BibitemShut {NoStop}%
\bibitem [{\citenamefont {Becattini}\ and\ \citenamefont {Rindori}(2019)}]{Becattini:2019poj}%
  \BibitemOpen
  \bibfield  {author} {\bibinfo {author} {\bibfnamefont {F.}~\bibnamefont {Becattini}}\ and\ \bibinfo {author} {\bibfnamefont {D.}~\bibnamefont {Rindori}},\ }\href {\doibase 10.1103/PhysRevD.99.125011} {\bibfield  {journal} {\bibinfo  {journal} {Phys. Rev. D}\ }\textbf {\bibinfo {volume} {99}},\ \bibinfo {pages} {125011} (\bibinfo {year} {2019})},\ \Eprint {http://arxiv.org/abs/1903.05422} {arXiv:1903.05422 [hep-th]} \BibitemShut {NoStop}%
\bibitem [{\citenamefont {Becattini}\ \emph {et~al.}(2021)\citenamefont {Becattini}, \citenamefont {Buzzegoli},\ and\ \citenamefont {Palermo}}]{Becattini:2020qol}%
  \BibitemOpen
  \bibfield  {author} {\bibinfo {author} {\bibfnamefont {F.}~\bibnamefont {Becattini}}, \bibinfo {author} {\bibfnamefont {M.}~\bibnamefont {Buzzegoli}}, \ and\ \bibinfo {author} {\bibfnamefont {A.}~\bibnamefont {Palermo}},\ }\href {\doibase 10.1007/JHEP02(2021)101} {\bibfield  {journal} {\bibinfo  {journal} {JHEP}\ }\textbf {\bibinfo {volume} {02}},\ \bibinfo {pages} {101} (\bibinfo {year} {2021})},\ \Eprint {http://arxiv.org/abs/2007.08249} {arXiv:2007.08249 [hep-th]} \BibitemShut {NoStop}%
\bibitem [{\citenamefont {Palermo}\ \emph {et~al.}(2021)\citenamefont {Palermo}, \citenamefont {Buzzegoli},\ and\ \citenamefont {Becattini}}]{Palermo:2021hlf}%
  \BibitemOpen
  \bibfield  {author} {\bibinfo {author} {\bibfnamefont {A.}~\bibnamefont {Palermo}}, \bibinfo {author} {\bibfnamefont {M.}~\bibnamefont {Buzzegoli}}, \ and\ \bibinfo {author} {\bibfnamefont {F.}~\bibnamefont {Becattini}},\ }\href {\doibase 10.1007/JHEP10(2021)077} {\bibfield  {journal} {\bibinfo  {journal} {JHEP}\ }\textbf {\bibinfo {volume} {10}},\ \bibinfo {pages} {077} (\bibinfo {year} {2021})},\ \Eprint {http://arxiv.org/abs/2106.08340} {arXiv:2106.08340 [hep-th]} \BibitemShut {NoStop}%
\bibitem [{\citenamefont {Gavassino}(2020)}]{GavassinoTermometri}%
  \BibitemOpen
  \bibfield  {author} {\bibinfo {author} {\bibfnamefont {L.}~\bibnamefont {Gavassino}},\ }\href {\doibase 10.1007/s10701-020-00393-x} {\bibfield  {journal} {\bibinfo  {journal} {Found. Phys.}\ }\textbf {\bibinfo {volume} {50}},\ \bibinfo {pages} {1554} (\bibinfo {year} {2020})},\ \Eprint {http://arxiv.org/abs/2005.06396} {arXiv:2005.06396 [gr-qc]} \BibitemShut {NoStop}%
\bibitem [{\citenamefont {{Gavassino}}(2021)}]{GavassinoGibbs2021}%
  \BibitemOpen
  \bibfield  {author} {\bibinfo {author} {\bibfnamefont {L.}~\bibnamefont {{Gavassino}}},\ }\href {\doibase 10.1088/1361-6382/ac2b0e} {\bibfield  {journal} {\bibinfo  {journal} {Classical and Quantum Gravity}\ }\textbf {\bibinfo {volume} {38}},\ \bibinfo {eid} {21LT02} (\bibinfo {year} {2021})},\ \Eprint {http://arxiv.org/abs/2104.09142} {arXiv:2104.09142 [gr-qc]} \BibitemShut {NoStop}%
\bibitem [{\citenamefont {Gavassino}\ \emph {et~al.}(2022)\citenamefont {Gavassino}, \citenamefont {Antonelli},\ and\ \citenamefont {Haskell}}]{GavassinoCausality2021}%
  \BibitemOpen
  \bibfield  {author} {\bibinfo {author} {\bibfnamefont {L.}~\bibnamefont {Gavassino}}, \bibinfo {author} {\bibfnamefont {M.}~\bibnamefont {Antonelli}}, \ and\ \bibinfo {author} {\bibfnamefont {B.}~\bibnamefont {Haskell}},\ }\href {\doibase 10.1103/PhysRevLett.128.010606} {\bibfield  {journal} {\bibinfo  {journal} {Phys. Rev. Lett.}\ }\textbf {\bibinfo {volume} {128}},\ \bibinfo {pages} {010606} (\bibinfo {year} {2022})},\ \Eprint {http://arxiv.org/abs/2105.14621} {arXiv:2105.14621 [gr-qc]} \BibitemShut {NoStop}%
\bibitem [{\citenamefont {Gavassino}(2023)}]{GavassinoSymmetric2022nff}%
  \BibitemOpen
  \bibfield  {author} {\bibinfo {author} {\bibfnamefont {L.}~\bibnamefont {Gavassino}},\ }\href {\doibase 10.1103/PhysRevD.107.065013} {\bibfield  {journal} {\bibinfo  {journal} {Phys. Rev. D}\ }\textbf {\bibinfo {volume} {107}},\ \bibinfo {pages} {065013} (\bibinfo {year} {2023})},\ \Eprint {http://arxiv.org/abs/2210.05067} {arXiv:2210.05067 [nucl-th]} \BibitemShut {NoStop}%
\bibitem [{\citenamefont {Mullins}\ \emph {et~al.}(2023)\citenamefont {Mullins}, \citenamefont {Hippert},\ and\ \citenamefont {Noronha}}]{MullinsInfo2023tjg}%
  \BibitemOpen
  \bibfield  {author} {\bibinfo {author} {\bibfnamefont {N.}~\bibnamefont {Mullins}}, \bibinfo {author} {\bibfnamefont {M.}~\bibnamefont {Hippert}}, \ and\ \bibinfo {author} {\bibfnamefont {J.}~\bibnamefont {Noronha}},\ }\href {\doibase 10.1103/PhysRevD.108.076013} {\bibfield  {journal} {\bibinfo  {journal} {Phys. Rev. D}\ }\textbf {\bibinfo {volume} {108}},\ \bibinfo {pages} {076013} (\bibinfo {year} {2023})},\ \Eprint {http://arxiv.org/abs/2306.08635} {arXiv:2306.08635 [nucl-th]} \BibitemShut {NoStop}%
\bibitem [{\citenamefont {Gavassino}\ \emph {et~al.}(2024)\citenamefont {Gavassino}, \citenamefont {Disconzi},\ and\ \citenamefont {Noronha}}]{GavassinoUniveraalityI2023odx}%
  \BibitemOpen
  \bibfield  {author} {\bibinfo {author} {\bibfnamefont {L.}~\bibnamefont {Gavassino}}, \bibinfo {author} {\bibfnamefont {M.~M.}\ \bibnamefont {Disconzi}}, \ and\ \bibinfo {author} {\bibfnamefont {J.}~\bibnamefont {Noronha}},\ }\href {\doibase 10.1103/PhysRevLett.132.222302} {\bibfield  {journal} {\bibinfo  {journal} {Phys. Rev. Lett.}\ }\textbf {\bibinfo {volume} {132}},\ \bibinfo {pages} {222302} (\bibinfo {year} {2024})},\ \Eprint {http://arxiv.org/abs/2302.03478} {arXiv:2302.03478 [nucl-th]} \BibitemShut {NoStop}%
\bibitem [{\citenamefont {Mullins}\ \emph {et~al.}(2025)\citenamefont {Mullins}, \citenamefont {Hippert},\ and\ \citenamefont {Noronha}}]{Mullins:2025vqa}%
  \BibitemOpen
  \bibfield  {author} {\bibinfo {author} {\bibfnamefont {N.}~\bibnamefont {Mullins}}, \bibinfo {author} {\bibfnamefont {M.}~\bibnamefont {Hippert}}, \ and\ \bibinfo {author} {\bibfnamefont {J.}~\bibnamefont {Noronha}},\ }\href {\doibase 10.1103/PhysRevLett.134.232302} {\bibfield  {journal} {\bibinfo  {journal} {Phys. Rev. Lett.}\ }\textbf {\bibinfo {volume} {134}},\ \bibinfo {pages} {232302} (\bibinfo {year} {2025})},\ \Eprint {http://arxiv.org/abs/2501.04637} {arXiv:2501.04637 [nucl-th]} \BibitemShut {NoStop}%
\bibitem [{\citenamefont {{Pei}}\ \emph {et~al.}(2025)\citenamefont {{Pei}}, \citenamefont {{Chen}},\ and\ \citenamefont {{Quan}}}]{Pei2025}%
  \BibitemOpen
  \bibfield  {author} {\bibinfo {author} {\bibfnamefont {J.-H.}\ \bibnamefont {{Pei}}}, \bibinfo {author} {\bibfnamefont {J.-F.}\ \bibnamefont {{Chen}}}, \ and\ \bibinfo {author} {\bibfnamefont {H.~T.}\ \bibnamefont {{Quan}}},\ }\href {\doibase 10.1103/xlmq-g6m5} {\bibfield  {journal} {\bibinfo  {journal} {\prl}\ }\textbf {\bibinfo {volume} {134}},\ \bibinfo {eid} {237102} (\bibinfo {year} {2025})},\ \Eprint {http://arxiv.org/abs/2312.17621} {arXiv:2312.17621 [cond-mat.stat-mech]} \BibitemShut {NoStop}%
\bibitem [{\citenamefont {Mizuno}\ and\ \citenamefont {Rezzolla}(2025)}]{Mizuno:2024tis}%
  \BibitemOpen
  \bibfield  {author} {\bibinfo {author} {\bibfnamefont {Y.}~\bibnamefont {Mizuno}}\ and\ \bibinfo {author} {\bibfnamefont {L.}~\bibnamefont {Rezzolla}},\ }\enquote {\bibinfo {title} {{General-Relativistic Magnetohydrodynamic Equations: The Bare Essential}},}\ \ (\bibinfo {year} {2025})\ \Eprint {http://arxiv.org/abs/2404.13824} {arXiv:2404.13824 [astro-ph.HE]} \BibitemShut {NoStop}%
\bibitem [{\citenamefont {Tolman}\ and\ \citenamefont {Stewart}(1916)}]{TolmanStewart1916}%
  \BibitemOpen
  \bibfield  {author} {\bibinfo {author} {\bibfnamefont {R.~C.}\ \bibnamefont {Tolman}}\ and\ \bibinfo {author} {\bibfnamefont {T.~D.}\ \bibnamefont {Stewart}},\ }\href {\doibase 10.1103/PhysRev.8.97} {\bibfield  {journal} {\bibinfo  {journal} {Phys. Rev.}\ }\textbf {\bibinfo {volume} {8}},\ \bibinfo {pages} {97} (\bibinfo {year} {1916})}\BibitemShut {NoStop}%
\bibitem [{\citenamefont {{Wang}}(2023)}]{WangStewartTolman2023}%
  \BibitemOpen
  \bibfield  {author} {\bibinfo {author} {\bibfnamefont {Z.-Y.}\ \bibnamefont {{Wang}}},\ }\href {\doibase 10.1007/s00339-022-06344-9} {\bibfield  {journal} {\bibinfo  {journal} {Applied Physics A: Materials Science \& Processing}\ }\textbf {\bibinfo {volume} {129}},\ \bibinfo {eid} {71} (\bibinfo {year} {2023})}\BibitemShut {NoStop}%
\bibitem [{\citenamefont {Bei}(2025)}]{BeiPhD}%
  \BibitemOpen
  \bibfield  {author} {\bibinfo {author} {\bibfnamefont {G.}~\bibnamefont {Bei}},\ }\emph {\bibinfo {title} {Diffusione anisotropa e ondulatoria del calore e effetti termomagnetici e termoelettrici oscillanti autoindotti sui conduttori rotanti.}},\ \href@noop {} {Ph.D. thesis},\ \bibinfo  {school} {Universit\'{a} di Roma Sapienza} (\bibinfo {year} {2025})\BibitemShut {NoStop}%
\bibitem [{\citenamefont {Bellan}(2006)}]{bellan_2006}%
  \BibitemOpen
  \bibfield  {author} {\bibinfo {author} {\bibfnamefont {P.~M.}\ \bibnamefont {Bellan}},\ }\href {\doibase 10.1017/CBO9780511807183} {\emph {\bibinfo {title} {Fundamentals of Plasma Physics}}}\ (\bibinfo  {publisher} {Cambridge University Press},\ \bibinfo {year} {2006})\BibitemShut {NoStop}%
\bibitem [{\citenamefont {Huang}(1987)}]{huang_book}%
  \BibitemOpen
  \bibfield  {author} {\bibinfo {author} {\bibfnamefont {K.}~\bibnamefont {Huang}},\ }\href@noop {} {\emph {\bibinfo {title} {Statistical Mechanics}}},\ \bibinfo {edition} {2nd}\ ed.\ (\bibinfo  {publisher} {John Wiley \& Sons},\ \bibinfo {year} {1987})\BibitemShut {NoStop}%
\bibitem [{\citenamefont {Srednicki}(2007)}]{Srednicki_2007}%
  \BibitemOpen
  \bibfield  {author} {\bibinfo {author} {\bibfnamefont {M.}~\bibnamefont {Srednicki}},\ }\href@noop {} {\emph {\bibinfo {title} {Quantum Field Theory}}}\ (\bibinfo  {publisher} {Cambridge University Press},\ \bibinfo {year} {2007})\BibitemShut {NoStop}%
\bibitem [{\citenamefont {Hernandez}\ and\ \citenamefont {Kovtun}(2017)}]{Hernandez:2017mch}%
  \BibitemOpen
  \bibfield  {author} {\bibinfo {author} {\bibfnamefont {J.}~\bibnamefont {Hernandez}}\ and\ \bibinfo {author} {\bibfnamefont {P.}~\bibnamefont {Kovtun}},\ }\href {\doibase 10.1007/JHEP05(2017)001} {\bibfield  {journal} {\bibinfo  {journal} {JHEP}\ }\textbf {\bibinfo {volume} {05}},\ \bibinfo {pages} {001} (\bibinfo {year} {2017})},\ \Eprint {http://arxiv.org/abs/1703.08757} {arXiv:1703.08757 [hep-th]} \BibitemShut {NoStop}%
\bibitem [{\citenamefont {Gavassino}\ and\ \citenamefont {Shokri}(2023)}]{GavassinoShokryMHD:2023qnw}%
  \BibitemOpen
  \bibfield  {author} {\bibinfo {author} {\bibfnamefont {L.}~\bibnamefont {Gavassino}}\ and\ \bibinfo {author} {\bibfnamefont {M.}~\bibnamefont {Shokri}},\ }\href {\doibase 10.1103/PhysRevD.108.096010} {\bibfield  {journal} {\bibinfo  {journal} {Phys. Rev. D}\ }\textbf {\bibinfo {volume} {108}},\ \bibinfo {pages} {096010} (\bibinfo {year} {2023})},\ \Eprint {http://arxiv.org/abs/2307.11615} {arXiv:2307.11615 [nucl-th]} \BibitemShut {NoStop}%
\bibitem [{\citenamefont {Landau}\ and\ \citenamefont {Lifshitz}(2013)}]{landau5}%
  \BibitemOpen
  \bibfield  {author} {\bibinfo {author} {\bibfnamefont {L.}~\bibnamefont {Landau}}\ and\ \bibinfo {author} {\bibfnamefont {E.}~\bibnamefont {Lifshitz}},\ }\href {https://books.google.pl/books?id=VzgJN-XPTRsC} {\emph {\bibinfo {title} {Statistical Physics}}},\ \bibinfo {number} {v. 5}\ (\bibinfo  {publisher} {Elsevier Science},\ \bibinfo {year} {2013})\BibitemShut {NoStop}%
\bibitem [{\citenamefont {Weinberg}(1995)}]{weinbergQFT_1995}%
  \BibitemOpen
  \bibfield  {author} {\bibinfo {author} {\bibfnamefont {S.}~\bibnamefont {Weinberg}},\ }\href {\doibase 10.1017/CBO9781139644167} {\emph {\bibinfo {title} {The Quantum Theory of Fields}}},\ Vol.~\bibinfo {volume} {1}\ (\bibinfo  {publisher} {Cambridge University Press},\ \bibinfo {year} {1995})\BibitemShut {NoStop}%
\bibitem [{\citenamefont {Wald}(1984)}]{Wald}%
  \BibitemOpen
  \bibfield  {author} {\bibinfo {author} {\bibfnamefont {R.~M.}\ \bibnamefont {Wald}},\ }\href {https://cds.cern.ch/record/106274} {\emph {\bibinfo {title} {{General relativity}}}}\ (\bibinfo  {publisher} {Chicago Univ. Press},\ \bibinfo {address} {Chicago, IL},\ \bibinfo {year} {1984})\BibitemShut {NoStop}%
\bibitem [{\citenamefont {{Rezzolla}}\ and\ \citenamefont {{Zanotti}}(2013)}]{rezzolla_book}%
  \BibitemOpen
  \bibfield  {author} {\bibinfo {author} {\bibfnamefont {L.}~\bibnamefont {{Rezzolla}}}\ and\ \bibinfo {author} {\bibfnamefont {O.}~\bibnamefont {{Zanotti}}},\ }\href@noop {} {\emph {\bibinfo {title} {Relativistic Hydrodynamics, by L.~Rezzolla and O.~Zanotti.~Oxford University Press, 2013.~ISBN-10: 0198528906; ISBN-13: 978-0198528906}}}\ (\bibinfo {year} {2013})\BibitemShut {NoStop}%
\bibitem [{\citenamefont {{Born}}(1909)}]{Born1909}%
  \BibitemOpen
  \bibfield  {author} {\bibinfo {author} {\bibfnamefont {M.}~\bibnamefont {{Born}}},\ }\href {\doibase 10.1002/andp.19093351102} {\bibfield  {journal} {\bibinfo  {journal} {Annalen der Physik}\ }\textbf {\bibinfo {volume} {335}},\ \bibinfo {pages} {1} (\bibinfo {year} {1909})}\BibitemShut {NoStop}%
\bibitem [{\citenamefont {{Cercignani}}\ and\ \citenamefont {{Kremer}}(2002)}]{cercignani_book}%
  \BibitemOpen
  \bibfield  {author} {\bibinfo {author} {\bibfnamefont {C.}~\bibnamefont {{Cercignani}}}\ and\ \bibinfo {author} {\bibfnamefont {G.~M.}\ \bibnamefont {{Kremer}}},\ }\href@noop {} {\emph {\bibinfo {title} {{The relativistic Boltzmann equation: theory and applications}}}}\ (\bibinfo {year} {2002})\BibitemShut {NoStop}%
\bibitem [{\citenamefont {de~Groot}\ \emph {et~al.}(1980)\citenamefont {de~Groot}, \citenamefont {van Leeuwen},\ and\ \citenamefont {van Weert}}]{Groot1980RelativisticKT}%
  \BibitemOpen
  \bibfield  {author} {\bibinfo {author} {\bibfnamefont {S.~R.}\ \bibnamefont {de~Groot}}, \bibinfo {author} {\bibfnamefont {W.~A.}\ \bibnamefont {van Leeuwen}}, \ and\ \bibinfo {author} {\bibfnamefont {C.~G.}\ \bibnamefont {van Weert}},\ }\href@noop {} {\emph {\bibinfo {title} {Relativistic kinetic theory: principles and applications}}}\ (\bibinfo {year} {1980})\BibitemShut {NoStop}%
\bibitem [{\citenamefont {Poisson}(2009)}]{PoissonToolkit2009pwt}%
  \BibitemOpen
  \bibfield  {author} {\bibinfo {author} {\bibfnamefont {E.}~\bibnamefont {Poisson}},\ }\href {\doibase 10.1017/CBO9780511606601} {\emph {\bibinfo {title} {{A Relativist's Toolkit: The Mathematics of Black-Hole Mechanics}}}}\ (\bibinfo  {publisher} {Cambridge University Press},\ \bibinfo {year} {2009})\BibitemShut {NoStop}%
\bibitem [{\citenamefont {Gavassino}(2022)}]{GavassinoCovarianceThermo:2021kia}%
  \BibitemOpen
  \bibfield  {author} {\bibinfo {author} {\bibfnamefont {L.}~\bibnamefont {Gavassino}},\ }\href {\doibase 10.1007/s10701-021-00518-w} {\bibfield  {journal} {\bibinfo  {journal} {Found. Phys.}\ }\textbf {\bibinfo {volume} {52}},\ \bibinfo {pages} {11} (\bibinfo {year} {2022})},\ \Eprint {http://arxiv.org/abs/2105.09294} {arXiv:2105.09294 [gr-qc]} \BibitemShut {NoStop}%
\end{thebibliography}%

\label{lastpage}

\end{document}